\documentclass[prd,aps,twocolumn,reprint,groupedaddress,superscriptaddress,nofootinbib]{revtex4-1}
\usepackage{graphicx}
\usepackage{amsmath,amssymb,bm}
\usepackage{hyperref}
\usepackage{autobreak,mathtools}
\usepackage{comment}
\usepackage{mathrsfs}
\usepackage[dvipsnames]{xcolor}
\usepackage[normalem]{ulem}
\usepackage{colortbl}

\definecolor{p-r}{RGB}{210, 110, 100}
\hypersetup{colorlinks=true, citecolor=p-r, linkcolor=p-r,urlcolor=p-r}

\begin{document}

\title{Regular Vaidya Solutions of Effective Gravitational Theories}
\author{Valentin Boyanov}
\affiliation{CENTRA, Departamento de F\'isica, Instituto Superior T\'ecnico - IST, Universidade de Lisboa - UL, Avenida Rovisco Pais 1, 1049 Lisboa, Portugal}

\author{Ra\'ul Carballo-Rubio}
\affiliation{Instituto de Astrof\'isica de Andaluc\'ia (IAA-CSIC),
Glorieta de la Astronom\'ia, 18008 Granada, Spain}
\affiliation{Center of Gravity, Niels Bohr Institute, Blegdamsvej 17, 2100 Copenhagen, Denmark}

\begin{abstract}
We initiate the study of the dynamics of spherically symmetric spacetimes beyond general relativity through exact solutions of the field equations of second-order effective gravitational theories defined solely in terms of the symmetries of the problem in four or more dimensions. We prove the existence of regular Vaidya solutions that represent the formation or change in mass of regular black holes (or black hole mimickers) due to the gravitational collapse or accretion of radiation, with the time reverse of these processes being also included as solutions. Our treatment and results are remarkable in their simplicity and the breadth of possibilities they provide for novel theoretical applications and discoveries in the context of regular gravitational collapse. For instance, we illustrate how modifications of the gravitational dynamics can cure curvature singularities through what can be interpreted as an energy transfer between matter and gravitational degrees of freedom. As another application, we show how to incorporate spacetimes frequently used in the literature, designed geometrically to represent the formation and disappearance of regular black holes, as fully dynamical solutions within this formalism.
\end{abstract}

\maketitle

\textit{Introduction---}Black holes are one of the most fascinating theoretical constructs in gravitational physics. Each of their characteristic elements is an example of extreme behaviors of spacetime curvature due to gravitational effects: photon spheres illustrate the extreme bending of light into bounded orbits~\cite{Luminet:1979nyg,Perlick:2021aok}, event horizons demonstrate the eternal capture of matter and energy inside black hole interiors~\cite{Rindler:1956yx,Gourgoulhon:2008pu}, and singularities indicate the unbounded growth of gravitational fields and the consequent breakdown of general relativity itself~\cite{Penrose:1964wq,Senovilla:2014gza}.

While observations have thus far shown an excellent agreement with the predictions of general relativity for the spacetime in the exterior of these gravitational objects~\cite{LIGOScientific:2016aoc,Cahillane:2022pqm,EventHorizonTelescope:2019dse,EventHorizonTelescope:2022wkp}, numerous conceptual difficulties plague the description of their interior; namely, the aforementioned existence of singularities~\cite{Penrose:1964wq,Senovilla:2014gza}, the information loss problem~\cite{Unruh:2017uaw,Buoninfante:2021ijy}, and the instability of inner horizons~\cite{Simpson:1973ua,Poisson1989,Poisson1990,Ori:1991zz}.

It is widely expected that a full solution of these difficulties requires unifying quantum field theory and general relativity. While existing theories of quantum gravity are still far from providing a complete picture~\cite{Carballo-Rubio:2025fnc}, the leading low-energy effects of such theories can often be described in terms of effective geometric modifications of general relativity (recent reviews of gravitational collapse in quantum gravity include~\cite{Eichhorn:2022bgu,Ashtekar:2023cod,Platania:2023srt}). Research along this line has, however, stumbled upon substantial difficulties. There is evidence that regularizing general black hole interiors requires field equations of order higher than second~\cite{Knorr:2022kqp,Borissova:2023kzq}, but higher-order field equations pose important challenges even for basic working notions such as well posedness~\cite{Kovacs:2020pns,Kovacs:2020ywu,Brady:2023dgu}.

This Letter presents a breakthrough in the study of the effective theories describing black holes. We show for the first time the existence of simple exact solutions representing the formation of regular black holes for a wide class of second-order gravitational theories in 4 or more dimensions. The starting point of the analysis is the definition of the space of second-order theories compatible with the symmetries of the problem, with no further restrictions. We then select the matter source to search for Vaidya solutions in this general space, and determine the subset of these theories for which these solutions are singularity-free.

In general relativity, the Vaidya solution is a generalization of the Schwarzschild solution with null dust (radiation) as a matter source that modifies the mass of the central black hole~\cite{Vaidya:1951fdr,Vaidya:1953zza,Vaidya:1966zza}, or its time reversal in which the central object is instead a white hole. Because of its simplicity and physical significance, this key solution is used extensively in discussions of black hole physics across different levels, from textbooks and lecture notes on the subject~\cite{Strominger:1994tn,Poisson:2009pwt} to research papers discussing core aspects of black hole physics, including diverse definitions of black hole boundaries~\cite{Schnetter:2005ea,Ben-Dov:2006zmw,Bengtsson:2008jr,Nielsen:2010wq}, the phenomenon of mass inflation~\cite{Ori:1991zz,Frolov:2017rjz,Bonanno:2020fgp,Carballo-Rubio:2021bpr,Carballo-Rubio:2024dca}, and black hole evaporation~\cite{Parikh:1998ux,Frolov:2014jva,Hayward:2005gi}. Showing that an equally simple generalization of this famous solution exists for second-order effective gravitational theories, including cases describing dynamical regular black holes instead, is a crucial development in the study of physics beyond general relativity.

Our results also present a rich and interesting interplay with other developments. A key takeaway of this Letter is that an effective geometric description captures the main ingredients of the dynamical formation of regular black holes in spherical symmetry, including and complementing the study of solutions of specific gravitational theories defined in five or higher dimensions~\cite{Bueno:2024eig}, while allowing us both to be more general and construct solutions in the more interesting case of four dimensions. One concrete application, described below, of our formalism is promoting kinematically prescribed regular black hole geometries, such as the one introduced in the influential work of Hayward~\cite{Hayward:2005gi}, to dynamical solutions. Our formalism also provides a natural framework to discuss theories with a varying gravitational coupling, and the connection with specific implementations, such as the one recently discussed in the context of the gravitational collapse of timelike dust~\cite{Bonanno:2023rzk}, is worth exploring in detail.

\textit{Second-order effective gravitational actions---}In the following, we focus on spherically symmetric situations. While our main aim is working in $D=4$ dimensions, the formalism introduced below can describe in a unified manner the dynamics of spacetimes for any dimensionality $D\geq 4$. Keeping the dimensionality general will also be useful in the discussion of some of the examples below. We can thus consider a general metric of the form
\begin{equation}\label{eq:Dsymred}
g^{(D)}_{\mu\nu}(y)\text{d}y^\mu\text{d}y^\nu=g_{ab}(x)\text{d}x^a\text{d}x^b+R^2(x)\text{d}\Omega_{(D-2)}^2,
\end{equation}
where $D$-dimensional coordinates are labeled by $\{y^\mu\}_{\mu=0}^{D-1}$, $R(x)$ is a scalar field on the two-dimensional space with coordinates $\{x^a=x^0,x^1\}$ (and will itself be used as one of these coordinates later on), and $\text{d}\Omega_{(D-2)}^2$ is the line element on the unit $(D-2)$-sphere.

We start considering as motivation the dynamics of the $D$-dimensional metric $g^{(D)}_{\mu\nu}$ and minimally coupled matter fields, described by a $D$-dimensional action $\mathscr{S}^{(D)}=\int\text{d}^Dy\sqrt{-g^{(D)}}\left[\mathscr{L}_{\rm G}^{(D)}/\kappa+\mathscr{L}_{\rm M}^{(D)}\right]$, where $\mathscr{L}_{\rm G}^{(D)}$ and $\mathscr{L}_{\rm M}^{(D)}$ are the gravitational and matter Lagrangian densities, respectively, and $\kappa=16\pi$ in natural units. Palais's principle of symmetric criticality~\cite{Palais:1979rca,Fels:2001rv,Frausto:2024egp} is guaranteed to hold due to the rotation Lie group being compact (see, e.g., proposition 5.11 in~\cite{Fels:2001rv}), and therefore performing variations with respect to $g^{(D)}_{\mu\nu}(y)$ and restricting the field equations to spherically symmetric metrics of the form~\eqref{eq:Dsymred} is equivalent to restricting the $D$-dimensional action to spherically symmetric metrics and performing variations with respect to $g_{ab}(x)$ and $R(x)$.

When restricting the action $\mathscr{S}^{(D)}$ to spherically symmetric metrics, the resulting gravitational Lagrangian is $\Omega_{(D-2)}\mathscr{L}_{\rm G}$, where $\Omega_{(D-2)}=\int\text{d}\Omega_{(D-2)}=2\pi^{(D-1)/2}/\Gamma(D/2-1/2)$  is a numerical constant and $\mathscr{L}_{\rm G}$ is a function of a two-dimensional metric $g_{ab}(x)$ and a scalar field $R(x)$. In general, the equations of motion will be of arbitrarily high order, depending on the original $D$-dimensional theory considered. This space of theories is completely general and includes the dynamics of the spherically symmetric sector of any $D$-dimensional metric theory that admits a variational principle formulation.

Within this general space of theories, there exists a subset with second-order field equations. This is a corollary of the seminal work of Horndeski (see, e.g.,~\cite{Horndeski:1974wa,Kobayashi:2019hrl}), particularized to a two-dimensional metric $g_{ab}$, which implies that the most general Lagrangian density satisfying this requirement can be written as
\begin{align}\label{eq:horact}
\mathscr{L}_{\rm G}&=H_2(R,X)-H_3(R,X)\square R+H_4(R,X)\mathcal{R}\nonumber\\
&-2\partial_X H_{4}(R,X)\left[(\square R)^2-\nabla^a\nabla^b R\nabla_a\nabla_b R\right],
\end{align}
where $\mathcal{R}$ is the Ricci scalar of $g_{ab}$, $\{H_i(R,X)\}_{i=2}^4$ are generic functions of two variables, and we have defined $X=\left(\nabla R\right)^2$. Let us note that, in higher dimensions, the most general Horndeski Lagrangian contains terms proportional to an additional function $H_5(R,X)$, but these terms vanish identically for two-dimensional metrics.

We can thus use two-dimensional Horndeski theories to describe the gravitational dynamics of $D$-dimensional spherically symmetric spacetimes (or other symmetric spacetimes that admit an equivalent reduction), exploiting the symmetries of the problem instead of committing to specific gravitational theories, which allows us to prove general results and distinguishes our approach from recent works~(e.g., \cite{Bueno:2024eig,Bueno:2024zsx,Bueno:2025gjg,Fernandes:2025fnz}). The geometries encompassed in our framework represent the spherically symmetric sector of all metric gravitational theories for which the field equations in this sector are of second order. Known families of theories satisfying this requirement in $D\geq5$ are Lovelock gravities~\cite{Kobayashi:2005ch,Maeda:2005ci,Dominguez:2005rt,Ghosh:2008jca,Cai:2008mh}, which are always of second order, and quasitopological gravities~\cite{Myers:2010ru,Oliva:2010eb,Bueno:2016xff,Hennigar:2017ego,Ahmed:2017jod,Bueno:2019ycr,Bueno:2022res,Moreno:2023rfl,Bueno:2024dgm}, which are of second order in the spherically symmetric sector, but generally of higher order when no symmetries are assumed. Since the first version of this Letter was posted on arXiv, more recent works have further explored the landscape of such theories, finding examples that are nonequivalent to general relativity also in $D=4$~\cite{Bueno:2025zaj,Borissova:2026dlz}. Our formalism can also accommodate Vaidya geometries obtained in recent discussions of effective spacetimes incorporating quantum gravity corrections, regardless of whether these are singular or regular~\cite{Borissova:2022mgd,Delaporte:2024and}.\footnote{While the second-order field equations used here cannot be equivalent to the higher-order field equations obtained from effective actions~\cite{Barvinsky:1985an,Fradkin:1983nw}, some dynamical aspects of singularity regularization or weakening can still be captured within such a truncation.} While the discussion below will be focused on regular geometries, this wide range of applications is interesting in its own right.

The spherically symmetric reduction of general relativity without a cosmological constant is recovered for $H_2=(D-2)(D-3)(1-X)R^{D-4}$, $H_3=2(D-2)R^{D-3}$ and $H_4=R^{D-2}$ (note that equivalence classes of these functions result in the same Lagrangian up to a boundary term). We will only consider theories that asymptote to general relativity at large distances, recovering these expressions in the $R\rightarrow\infty$ (weak gravity) limit.

Expressions for the variations of the Lagrangian density in Eq.~\eqref{eq:horact} are known~(see the detailed discussion in~\cite{Carballo-Rubio:2025ntd}). For our purposes here, it suffices to write explicitly the following three independent field equations in the presence of matter:
\begin{align}\label{eq:geneqs}
\beta\nabla_a\nabla_b R-g_{ab}\left(\alpha/2+\beta\square R \right)&\nonumber\\
+\left(\partial_X\alpha-\partial_R\beta \right)\nabla_aR\nabla_bR&=8\pi T_{ab},
\end{align}
where we have defined the two-dimensional source
\begin{equation}
T_{ab}=R^{D-2}T^{(D)}_{\mu\nu}\delta^\mu_a\delta^\nu_b,
\end{equation}
where the $D$-dimensional stress-energy tensor
$T^{(D)}_{\mu\nu}$ is defined using the standard Hilbert prescription, as well as the two functions
\begin{align}\label{eq:alphabetaH}
\alpha&=H_2+X\partial_R\left(H_3-2\partial_R H_4\right),\nonumber\\
\beta&=X\partial_X\left(H_3-2\partial_R H_4\right)-\partial_RH_4.
\end{align}
A specific representation of the functions $\{H_i\}_{i=2}^4$ for a given pair of functions $\alpha$ and $\beta$ is $H_2=\alpha$, $H_3=-2\beta$ and $H_4=-\int\text{d}R\,\beta$.

\textit{Regular Vaidya solutions---}In the rest of the Letter, we will be considering $D$-dimensional metrics of the form
\begin{equation}\label{eq:ansatz}
\text{d}s^2=-f(V,R)\text{d}V^2+2\text{d}V\text{d}R+R^2\text{d}\Omega^2_{D-2},
\end{equation}
where $f(V,R)$ is a function to be determined by solving the field equations, and $V$ is an ingoing null coordinate. Some geometric relations that will be useful later are $\nabla_V\nabla_VR=-\left(f\partial_R f-\partial_Vf\right)/2$, $\nabla_V\nabla_RR=\nabla_R\nabla_VR=\partial_Rf/2$,
$\nabla_R\nabla_RR=0$, $ (\nabla R)^2=f$ and $\square R=\partial_R f$. All calculations below apply, with just a few sign changes, to the time reversal of the metric~\eqref{eq:ansatz}, by implementing the transformation $V\rightarrow -V\propto U$, where $U$ would be an outgoing null coordinate.

The $D$-dimensional Vaidya solution~\cite{Iyer:1989nd} is a solution of the Einstein field equations with matter source
\begin{equation}\label{eq:vaidyasource}
T^{(D)}_{\mu\nu}=\frac{(D-2)\dot{M}(V)}{8\pi R^{D-2}} \partial_\mu V\partial_\nu V,
\end{equation}
where $\dot{M}(V)=\text{d}M(V)/\text{d}V$. 

We will now describe how the existence of solutions with this same matter source is a feature of theories beyond general relativity. The three independent components of Eq.~\eqref{eq:geneqs} yield the following equations:
\begin{align}
\partial_X\alpha-\partial_R\beta&=0,   \label{eq:rreqm}\\
\alpha+\beta \partial_Rf&=0,\label{eq:vreqm}\\
\beta\partial_Vf&=2(D-2)\dot{M}(V).\label{eq:vveq}   
\end{align}
Equation~\eqref{eq:rreqm} implies local exactness and hence the existence of a potential $\Theta\left(R,X\right)$ satisfying the off-shell relations
\begin{equation}\label{eq:solution2}
\partial_R\Theta(R,X)=\alpha(R,X),\quad \partial_X\Theta(R,X)=\beta(R,X), 
\end{equation}
Together with the remaining equations, the on-shell solution satisfies
\begin{equation}\label{eq:solution} \left.\Theta\left(R,X\right)\right|_{X=f(V,R)}=2(D-2)M(V).  
\end{equation}
Specific solutions can be generated in two ways, either by specifying functions $\alpha(R,X)$ and $\beta(R,X)$ satisfying the integrability condition~\eqref{eq:rreqm} and finding $\Theta(R,X)$ and $f(V,R)$, or by specifying a function $f(V,R)$ and deriving $\Theta(R,X)$ from which, using Eq.~\eqref{eq:solution2}, $\alpha(R,X)$ and $\beta(R,X)$ can be determined, and, in turn, the functions $\{H_i\}_{i=2}^4$ using the relations below Eq.~\eqref{eq:alphabetaH}. This algorithm can be used to construct a Lagrangian for any $\alpha(R,X)$ and $\beta(R,X)$ satisfying the integrability condition.

All Vaidya solutions discussed here are of type $D$ in the Petrov--Pirani--Penrose classification~\cite{Petrov:2000bs,Pirani:1956wr,Newman:1961qr,Penrose:1985bww}, as is the case in general relativity~\cite{Stephani:2003tm}. The principal double null directions $\{N^{(i)}\}_{i=1}^2$, satisfying $C_{\mu\nu\rho[\sigma}N^{(i)}_{\gamma]}N^{(i)\nu} N^{(i)\rho}=0$ where $C_{\mu\nu\rho\sigma}$ is the Weyl tensor, are the same as for general relativity (see, e.g.,~\cite{Coudray:2021pjr}), namely $N^{(1)}=\partial/\partial R$ and $N^{(2)}=\partial/\partial V+\left(f/2\right)\partial/\partial R$.

The degree of regularity of any geometry of the form in Eq.~\eqref{eq:Dsymred} can be checked by inspecting the components of curvature tensors in locally regular coordinate systems~\cite{Rinne:2005df} (see also ~\cite{Carballo-Rubio:2019fnb}). For the metric in Eq.~\eqref{eq:ansatz}, regularity in terms of finiteness of curvature scalars at the origin amounts to two conditions on the redshift function, $f|_{R=0}=1$ and $ \partial_Rf|_{R=0}=0$. Particularly, for the Kretschmann scalar the potentially divergent terms in $D\ge4$ would be proportional to $(f-1)^2/R^4$ and $(\partial_Rf)^2/R^2$, which is divergent unless these conditions are satisfied. The regularity conditions also imply, together with Eqs.~\eqref{eq:vreqm} and ~\eqref{eq:vveq}, the following on-shell divergent behavior in the $R\rightarrow0$ limit:
\begin{equation}\label{eq:regularityalphabeta}
\alpha(R,f)\propto R^p ,\quad \beta(R,f)\propto R^{p-1},\quad\text{with}\quad p\leq-1.
\end{equation}
The on-shell function $\Theta(R,f)$ does not depend on $R$, as shown in Eq.~\eqref{eq:solution}.

\textit{Examples---}We start noting that, for general relativity, $\alpha=(D-2)(D-3)R^{D-4}(1-X)$ and $\beta=-(D-2)R^{D-3}$. These functions satisfy Eq.~\eqref{eq:rreqm} and the corresponding Vaidya metric is generated by $\Theta(R,X)=(D-2)R^{D-3}(1-X)$, which results in $f(V,R)=1-2M(V)/R^{D-3}$. In all examples below, the asymptotic on-shell falloffs $f=1-2M(V)/R^{D-3}+\mathcal{O}( R^{2-D})$, $\alpha=2(D-2)(D-3)M(V)/R+\mathcal{O}(R^{-2})$ and $\beta=-2(D-2)R^{D-3}+\mathcal{O}(R^{D-4})$ guarantee recovering general relativity and asymptotic flatness at large distances.

The first example is the $D$-dimensional generalization of the theories defined for four-dimensional spherically symmetric spacetimes by Ziprick and Kunstatter~\cite{Ziprick:2010vb} (see also~\cite{Kunstatter:2015vxa,Barenboim:2024dko,Barenboim:2025ckx}), namely $H_2=(1-X)\text{d}^2\Phi(R)/\text{d}R^2$, $H_3=2\text{d}\Phi(R)/\text{d}R$ and $H_4=\Phi(R)$, with $\Phi(R)$ behaving as $\Phi(R)\simeq R^{D-2}$ for $R\rightarrow\infty$ (thus recovering general relativity at large distances). We thus have $\alpha=(1-X)\text{d}^2\Phi/\text{d}R^2$, $\beta=-\text{d}\Phi/dR$ which, in accordance with the regularity conditions discussed above, implies that $\Phi(R)$ must diverge at least as $\Phi(R)\propto R^{-1}$ for $R\rightarrow0$. We also have $\Theta(R,X)=(1-X)\text{d}\Phi/\text{d}R$ and $f(V,R)=1-2(D-2)M(V)/\left(\text{d}\Phi/\text{d}R\right)$. Solutions in this family are generalizations of the Bardeen spacetime~\cite{Bardeen:1968xts}, with the latter four-dimensional spacetime arising for $\Phi(R)=\left(R^2-2 \ell^2\right) \sqrt{R^2 + \ell^2}/(2R) +3\ell^2\arctan{\left(r/\sqrt{R^2+ \ell^2}\right)}/2$. We will refer to it as the Ziprick--Kunstatter (ZK) family.

The second example will be constructed following the inverse approach, starting from the $D$-dimensional generalization of the Hayward metric~\cite{Hayward:2005gi}, $f(V,R)=1-2R^2M(V)/\left[R^{D-1}+2\ell^{D-2}M(V)\right]$. Using this expression along with Eq.~\eqref{eq:solution}, we can determine that $\Theta(R,X)=(D-2)(1-X)R^{D-3}/\left[1-\ell^{D-2}(1-X)/R^2\right]$, from which $\alpha$ and $\beta$ (and the functions $\{H_i\}_{i=2}^4$) can be obtained. Building on this example, the ZK family can be generalised to the Ziprick--Kunstatter--Maeda--Taves (ZKMT) family $\Theta(R,X)=\Psi(Z)R^2\text{d}\Phi/\text{d}R$, where $\Psi(Z)$ is a function of the variable $Z=(1-X)/R^2$. The Hayward spacetime then corresponds to the particular choices $\Phi(R)=R^{D-2}$, $\Psi(Z)=Z/(1-\ell^{D-2}Z)$.

\begin{table*}
\begin{tabular}{|c|c|c|c|}
    \hline
    & General relativity & ZK & ZKMT \\
    \hline
    $\Theta(R,X)$ & $(D-2)(1-X)R^{D-3}$ & $(1-X)\text{d}\Phi(R)/\text{d}R$ & $\Psi(Z)R^2\text{d}\Phi(R)/\text{d}R$\\
    \hline
    $f(V,R)$ & $1-\frac{2M(V)}{R^{D-3}}$ & $1-\frac{2(D-2)M(V)}{\text{d}\Phi(R)/\text{d}R}$ & $1-R^2\Psi^{-1}\!\!\left[\frac{2(D-2)M}{R^2\text{d}\Phi(R)/\text{d}R}\right]$ \\
    \hline
    $f|_{R\sim 0}$ & $-\frac{2M(V)}{R^{D-3}}+\mathcal{O}(1)$ & $1-\frac{2(D-2)M(V)}{\left(R^{2}d\Phi/dR\right)|_{R=0}}R^2+\mathcal{O}(R^3)$ & $1-\left.\Psi^{-1}\!\!\left[\frac{2(D-2)M}{R^2\text{d}\Phi(R)/\text{d}R}\right]\right|_{R=0}R^2+\mathcal{O}(R^3)$ \\
    \hline
\end{tabular}
\caption{The potential function $\Theta$ for the example classes considered above. Regularity is assumed for the series expansions in the last two columns. We are considering the asymptotic on-shell falloffs $\Theta=2(D-2)M(V)+\mathcal{O}(R^{-1})$ and $f=1-2M(V)/R^{D-3}+\mathcal{O}(R^{2-D})$ for $R\rightarrow\infty$ in all examples. The functions $\alpha$ and $\beta$ can be obtained using Eq. \eqref{eq:solution2}.}
\label{tb:ex}
\end{table*}

All regular examples above illustrate that, for dimensional reasons, it is necessary to introduce at least one new length scale $\ell$ to ensure regularity. Depending on the relative value of $\ell$ (which we use as a placeholder for the potentially multiple new parameters in the solution) and $M(V)$ at a given moment of time, the corresponding geometry describes either a regular black hole or a horizonless object. Typically, for a given value of $\ell$, there is a mass threshold, determined kinematically, above which the corresponding geometries describe regular black holes~\cite{Carballo-Rubio:2022nuj}. For instance, for the Hayward example above in $D=4$, this threshold is $M=3\sqrt{3}\ell/4$~\cite{Hayward:2005gi}. For this threshold mass the function $f$ has a double root at $R=\sqrt{3}\ell$, making the geometry that of an extremal regular black hole. For masses above this threshold, the function $f$ has two roots that indicate the boundaries of the trapped region.

The value of the new parameter $\ell$ is not fixed \emph{a priori} in our approach. Choosing macroscopic values of $\ell$ can thus provide a phenomenological description of the formation of black hole mimickers (e.g., gravastarlike objects~\cite{Mazur:2001fv,Mazur:2004fk,Visser:2003ge}) from the collapse of null dust, as well as of radiation emission and absorption processes which change the mass of such objects. The study of black hole mimickers is an important area of research in its own right that has gained traction recently (see recent reviews in~\cite{Carballo-Rubio:2025fnc,Bambi:2025wjx}). Providing models for the dynamical formation and evolution of such objects is one of the key open challenges in this field, which our results can be directly applied to.

\textit{Singularity regularization and energy transfer---}The matter source in Eq.~\eqref{eq:vaidyasource} is singular at $R=0$. However, the corresponding Vaidya solutions are explicitly regular whenever the conditions in Eq.~\eqref{eq:regularityalphabeta} are satisfied.

A conceptual understanding of this issue is provided by the following splitting of the field equations:
\begin{align}\label{eq:effeinseq}
G^{(D)}_{\mu\nu}&=8\pi\left[ T^{(D)}_{\mu\nu}+\mathcal{T}^{(D)}_{\mu\nu}\right],
\end{align}
where $G^{(D)}_{\mu\nu}$ is the $D$-dimensional Einstein tensor, $T^{(D)}_{\mu\nu}$ is the matter source in Eq.~\eqref{eq:vaidyasource}, and $\mathcal{T}^{(D)}_{\mu\nu}$ is an effective stress-energy tensor coming from the terms regularizing the solutions to the Einstein equations. All these tensors are independently conserved.

For the regular Vaidya solutions, we have 
\begin{align}
G^{(D)}_{VV}&=-fG^{(D)}_{VR}-\frac{(D-2)^2}{R}\frac{\dot{M}}{\beta(R,f)},\nonumber\\  G^{(D)}_{VR}&=\frac{(D-2)(D-3)}{2R^2}\left(f-1\right)+\frac{(D-2)}{2R}\partial_Rf,\nonumber\\
8\pi\mathcal{T}^{(D)}_{VV}&=-fG^{(D)}_{VR}-\frac{(D-2)^2}{R}\frac{\dot{M}}{\beta(R,f)}-\frac{(D-2)\dot{M}}{R^{D-2}}.
\end{align}
These equations can be straightforwardly obtained by writing these components of the $D$-dimensional Einstein tensor in terms of variations of the gravitational action with respect to $g_{ab}$ and using Eq.~\eqref{eq:geneqs} for the particular case of general relativity, as well as Eq.~\eqref{eq:vveq}. All the components of the Einstein tensor are regular, while $\mathcal{T}^{(D)}_{VV}$ is singular and actually cancels the singularity in $T^{(D)}_{VV}$.

The only component of the matter stress-energy tensor $T^{(D)}_{\mu\nu}$ is an ingoing radial null flux $T^{(D)}_{VV}$ that is divergent at $R=0$, and its integral $\Omega_{(D-2)}T_{VV}=\Omega_{(D-2)}R^{D-2}T_{VV}^{(D)}$ is constant over all spheres throughout its collapse. On the other hand, the $\mathcal{T}^{(D)}_{VV}$ component of the effective stress-energy tensor contains an ingoing flux proportional to $\dot{M}$ that is also divergent at $R=0$, but with an overall integral over spheres that depends on the radius. Indeed, the piece proportional to $\dot{M}$ in $\Omega_{(D-2)}\mathcal{T}_{VV}=\Omega_{(D-2)}R^{D-2}\mathcal{T}_{VV}^{(D)}$ vanishes in the limit $R\rightarrow\infty$ and is switched on progressively as the radius decreases, until becoming $-\Omega_{(D-2)}T_{VV}$ exactly at $R=0$. Because of $\mathcal{T}^{(D)}_{\mu\nu}$ being conserved, this ingoing flux of energy must be created at the expense of leaving behind a residual distribution of energy, encoded in the remaining nonzero pieces of $\mathcal{T}^{(D)}_{\mu\nu}$, which describe how the energy in the collapsing or accreting null dust is stored in the gravitational field of the regular black hole after being effectively transferred.

\begin{figure}[h!]
    \centering
    \includegraphics[scale=0.9]{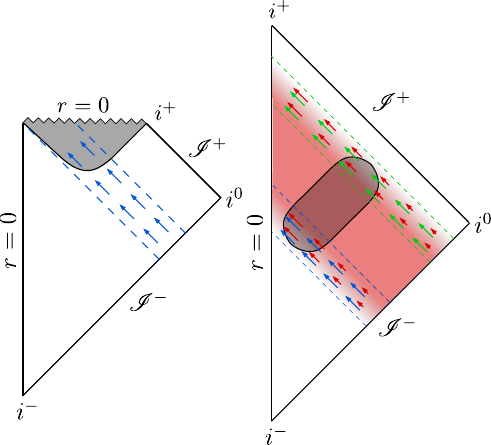}
\caption{\emph{Left:} Vaidya solution in general relativity representing the formation of a Schwarzschild black hole from the collapse of null dust originating from past null infinity, as encoded in the flux component of $T_{\mu\nu}^{(D)}$, indicated by blue arrows and contained within the blue dashed lines. The trapped region, characterized by $f(V,R)<0$, is shaded in gray. \emph{Right:} regular Vaidya solution representing the formation and disappearance of a regular black hole due to the collapse and accretion of null dust. Aside from the classical matter source $T^{(D)}_{\mu\nu}$, the effective source $\mathcal{T}^{(D)}_{\mu\nu}$ has two distinct components: an ingoing flux (red arrows), with energy of opposite sign to that of the null dust, and the energy distribution deposited in the regular black hole (red shaded region). The classical matter and effective sources cancel identically at $R=0$. The energy stored in the regular black hole grows while $\dot{M}>0$ (in between the blue dashed lines), decreases while $\dot{M}<0$ (in between the green dashed lines), and remains constant otherwise.
\label{fig:diagram}}
\end{figure}

The signs of these fluxes depend on the sign of $\dot{M}$. For $\dot{M}>0$ the flux in $T^{(D)}_{\mu\nu}$ has positive energy and the flux in $\mathcal{T}^{(D)}_{\mu\nu}$ has negative energy, while for $\dot{M}<0$ these signs are reversed. In other words, $\mathcal{T}^{(D)}_{\mu\nu}$ always acts as a regulator of $T^{(D)}_{\mu\nu}$ regardless of the sign of $\dot{M}(V)$. Depending on the evolution of $M(V)$, the corresponding geometries can describe processes of formation, accretion or disappearance of regular black holes. A direct calculation shows that the timelike convergence condition is necessarily violated by our regularity conditions, while the null convergence condition does not need to be violated (which is compatible with independently obtained results~\cite{Borissova:2025msp,Borissova:2025hmj}). As a consequence, the Hawking--Penrose singularity theorem~\cite{Hawking:1970zqf} does not hold, but the original Penrose theorem~\cite{Penrose:1964wq} may still appear applicable. However, a Cauchy horizon necessarily forms in all regular examples satisfying the null convergence condition, thus violating the requirement of global hyperbolicity necessary for the Penrose theorem. On the other hand, the formation of Cauchy horizons can be avoided if there exists a region where $\dot{M}<0$ (see Fig.~\ref{fig:diagram}), which, however, would violate the null convergence condition. We expect that similar observations would apply to formulations of singularity theorems with weakened energy conditions (e.g.,~\cite{Fewster:2010gm,Fewster:2019bjg}).

Combining both signs of $\dot{M}$ allows us to construct geometries with trapped regions but no event horizons, as depicted in Fig.~\ref{fig:diagram}. Such geometric constructions have frequently been used to model the expected effects of semiclassical backreaction, and are ubiquitous in the discussions of the evaporation of regular black holes, including~\cite{Hayward:2005gi,Frolov:2014jva,Frolov:2017rjz,Cardoso:2023guh,Carballo-Rubio:2024dca}. Here these can be constructed as exact solutions, which opens the possibility of a thorough understanding of the dynamical properties of evaporating regular black holes, such as their behavior under radial perturbations and the backreaction of the mass inflation instability~\cite{Frolov:2017rjz,Bonanno:2020fgp,Carballo-Rubio:2021bpr,Carballo-Rubio:2024dca}. Adding semiclassical sources, along the lines of the recent analysis for Reissner–Nordstr\"om black holes~\cite{Boyanov:2025otp} (see also~\cite{Barcelo:2020mjw,Barcelo:2022gii}), would also allow for a self-consistent modeling of the semiclassical dynamics of these regular objects. Recent related numerical breakthroughs can be found in~\cite{Barenboim:2024dko,Barenboim:2025ckx}.

\textit{Conclusions---}In this Letter, we have shown how the study of the field equations of general second-order effective gravitational theories describing spherically symmetric spacetimes can provide important insights into the geometric structure of black holes beyond general relativity. In particular, we have generalized the famous Vaidya solution of the Einstein-Hilbert theory to a wider class of gravitational actions, including theories in which the central curvature singularity is not formed. This result is remarkable due to its mathematical simplicity as well as its potential implications for the study of the effective low-energy dynamics of black holes in quantum gravity.

The singularity-free generalizations of the Vaidya solution discussed here maintain the simplicity that is characteristic of the original solution, while presenting new intriguing features that offer a novel window into the interplay of quantum effects and gravity. In particular, the energy in the collapsing null dust is transferred to gravitational degrees of freedom and stored in the resulting regular black hole. We hypothesize that this energy transfer may be a key aspect in the formation of macroscopic black holes in quantum gravity.

Another remarkable feature of our results is their generality. Our analysis of spherically symmetric geometries encompasses any new (quantum) gravitational effects describable with second-order field equations in $D\geq 4$ dimensions. To our knowledge, it is the first time that progress in the study of the dynamics of regular black holes has been made in such a general framework. While quantum gravity may ultimately involve higher-derivative field equations, the results discussed here are a strong indication that second-order field equations may suffice to describe key dynamical features of regular black holes and are thus worth exploring in detail.

Two particularly exciting potential applications of our results are the development of numerical codes that can handle the dynamical evolution of spherically symmetric regular black holes and the study of gravitational effects on matter and information beyond the test-field approximation in these extreme situations. Exact solutions such as the ones communicated here are important tools to provide insights into complex situations and validate numerical codes. Advancing along these lines has the potential of shedding new light onto long-standing issues such as the information loss problem and radically changing the way black holes are currently understood.

\acknowledgments

The authors are grateful to Jose Beltrán Jiménez for insightful discussions on two-dimensional Horndeski theories, Carlos Barcel\'o, Vitor Cardoso and Stefano Liberati for feedback on the first version of the manuscript, and \'Angel Murcia for useful correspondence. V.B. acknowledges support from the European Union’s H2020 ERC Advanced Grant “Black holes: gravitational engines of discovery” grant agreement no. Gravitas–101052587, as well as from the Spanish Government through the Grants No. PID2020-118159GB-C43, PID2020-118159GB-C44, PID2023-149018NB-C43 and PID2023-149018NB-C44 (funded by MCIN/AEI/10.13039/501100011033). R.C-R. acknowledges financial support provided by the Spanish Government through the Ram\'on y Cajal program (contract RYC2023-045894-I), the Grant No.~PID2023-149018NB-C43 funded~by MCIN/AEI/10.13039/501100011033, and the Severo Ochoa grant CEX2021-001131-S funded by MCIN/AEI/ 10.13039/501100011033, as well as the hospitality of the Center of Gravity, a Center of Excellence funded by the Danish National Research Foundation under grant No.~184.

\bibliographystyle{utphys}

\bibliography{refs}

\end{document}